\documentclass{revtex4}
\usepackage{epsfig}
\usepackage{amssymb}
\usepackage{graphicx}

\begin{document}
\title{Centrifugal quantum states of neutrons  }
\author{V.V. Nesvizhevsky}
 \author{A.K. Petukhov}
 \affiliation{Institut Laue-Langevin (ILL), 6 rue Jules Horowitz,
 F-38042, Grenoble, France, e-mail: nesvizhevsky@ill.eu}
 \author{K.V. Protasov}
 \affiliation { Laboratoire de Physique Subatomique et de Cosmologie
 (LPSC), IN2P3-CNRS, UJFG, 53, Avenue des Martyrs, F-38026, Grenoble,
 France}
 \author{A.Yu. Voronin}
 \affiliation {P.N. Lebedev Physical Institute, 53 Leninsky
 prospekt,119991, Moscow, Russia}

  \pacs{ 04.80.CC}

\begin{abstract}
We propose a method for observation of the quasi-stationary states of
neutrons, localized near the curved mirror surface. The bounding effective
well is formed by the
centrifugal potential and the mirror Fermi-potential.
This phenomenon is an example of an exactly solvable "quantum bouncer" problem
that could be studied experimentally.
It could provide a promising tool for studying fundamental neutron-matter interactions, as well as quantum neutron optics and surface physics effects.
We develop  formalism, which describes quantitatively   the
neutron motion near the mirror surface. The effects of mirror
roughness are taken into account.
\end{abstract}
\maketitle

\section{Introduction}
The "centrifugal states" of neutrons is a quantum analog of the so called whispering gallery wave, the phenomenon which in brief consists in  the wave localization near the curved  surface of a scatterer. It is known in acoustics since ancient times and was explained by Baron Rayleigh in his "Theory of Sound" \cite{Rayleigh,Ray1}. The whispering gallery waves in optics is an object of  growing interest during the last decade \cite{Oraev,Vahal}.
In the following we will be interested in the matter-wave aspect of the whispering gallery wave phenomenon, namely the large-angle neutron scattering on a curved mirror. Such a scattering can be explained in terms of states
of a quantum particle above a mirror in a linear potential - the so called "quantum bouncer"\cite{Gold, tHaar, LL, Flug, Sak, Langh, JGea}.
The neutron quantum motion in the Earth's gravitational field  above a flat mirror is another example of such a "quantum bouncer", which was observed  recently \cite{Nature1}.
We will show  that the centrifugal quantum bouncer and the gravitational quantum bouncer have many common features. Therefore we  compare these two phenomena and discuss  motivation for their studies.

 Experimental observation and study of the gravitational states is a challenging problem which  brings  rich physical information for  searches for extensions of the Standard model or for studying interaction of a quantum system with gravitational field \cite{Ahl,Khor,Bini,Kief,Lecl,AhlK,Ban,Brau,Boul,Buis,Acc,Saha,Man,Sil}, for constraining spin-independent extra short-range forces \cite{Bertol,NesPr,NesvPrP},  hypothetical axion-mediated spin-matter interactions \cite{BNPV}  and in surface physics.  Indeed any additional interaction between mirror bulk and neutron with the characteristic range of the gravitational states of a few micrometers would modify the quantum states and thus could be detected.

A natural extension of the mentioned experimental activity consists in approaching ultimate sensitivity for extra interactions at shorter characteristic ranges. Evidently, the quantum states characteristic size  has to be decreased. To achieve this goal  one needs to study novel approaches \cite{Wat}.  We will show that the promising method consists in localization of cold neutrons near a  curved mirror surface due to the superposition of the centrifugal potential and the Fermi potential of the mirror. In such a case the quasi-stationary "centrifugal" quantum states play essential role in the neutron flux dynamics. In the limit of  the centrifugal quantum states spatial size being much smaller than the curved mirror radius this problem is reduced to the simple case of a quantum particle in a linear  potential above a mirror.  Measurement of the gravitationally bound and centrifugal quantum states of neutrons could be considered as a kind of confirmation of the equivalence principle for a quantum particle \cite{Onof1,Onof2,Herd,Wawr,Chrys}. Both problems (the gravitational and the centrifugal ones) provide perfect experimental laboratory for studying neutron quantum optics phenomena, quantum revivals and localization \cite{Kalb, Rob,Berb, Bell, Math, Witt, Rom, Gonz}.  Evident advantages of using cold neutrons  consist in much higher statistics attainable, broad accessibility of cold neutron beams as well as in crucial reduction of many false effects compared to the experiments with the gravitationally bound quantum states of neutrons due to approximately $\sim 10^5$ times higher energies of the quantum states involved.

The phenomenon of the centrifugal quantum states of neutrons and the method of their experimental observation are described in Chapter 2. In Chapter 3 we develop  formalism, which describes neutron motion near the curved mirror surface; the properties of the centrifugal quantum states are discussed in Chapter 4. We will show that cold neutrons with the velocity of $\sim 10^3$ m/s  are well suited for such a kind of experiments. Time-dependent approach is considered in  Chapter 5. The effects of mirror roughness are taken into account in  Chapter 6.

 \section{Principle of observation}
If the neutron energy is much larger than the scatterer Fermi potential most neutrons are scattered to small angles. However some neutrons   could be captured into long-living centrifugal quasi-stationary states localized near the curved  scatterer surface and thus could be detected  at large deflection angles.
The curved mirror surface plays a role of a wave-guide and the centrifugal states play a role of radial modes in such a wave-guide. The spectral dependence of transmission probability  is determined by the existence of the centrifugal states in such a system.

 Similarly, in the gravitational state experiment one measures the slit-size dependence of the transmission probability of the wave-guide between a mirror and above-placed absorber \cite{Lusch1,Lusch2,Nesv1,Nesv2,PRD1,EPJC,PRD2,MeyNesv1,MeyNesv2,West}.
The characteristic energy scale of the gravitational state problem is  $\varepsilon_0=0.6$ $peV$, and the characteristic length scale is  $l_0=5.87$ $\mu m$. The mirror Fermi-potential could be considered as infinitely high and sharp. This approximation is justified as far as $l_0$  is much larger than the characteristic range of the Fermi-potential increase   (typically $<1$ nm), and $\varepsilon_0$  is much smaller than the characteristic value of the mirror Fermi-potential (typically $\sim 10^{-7}$ eV).
The methods for experimental observation of the gravitationally bound quantum states of neutrons are based on relatively large value of the characteristic length $l_0$, which allowed direct measurement of the shapes of neutron density distribution in the quantum states using two following complementary methods. First approach consisted in scanning the neutron density   above mirror using a flat horizontal scatterer/absorber at variable height. The second method is based on use of position-sensitive detectors of UCN with high spatial resolution of $\sim 1$ $\mu m$.

In analogy with the gravitational well the centrifugal quantum well is formed by effective centrifugal potential and repulsive Fermi-potential of a curved mirror as shown in Fig.1 and Fig.2. The effective acceleration near the curved mirror surface could be approximated as $a=v^2/R$ , where $v$  is the neutron velocity and $R$  is the mirror radius.
 We have significant freedom to choose  values of $v$  and $R$. In particular it would be advantageous to increase the neutron velocity and to decrease the mirror radius in order to get higher centrifugal acceleration $a$. In such a case the quantum well that confines the radial motion of neutrons near the  curved mirror surface   becomes  narrower, while the energy of radial motion in the corresponding quantum states increases. This enables us to eliminate many possible systematic effects. The radial  motion energy could be as high as the mirror Fermi-potential. In this case, we could use the Fermi-potential of a mirror as a "filter" for the quantum states.  For an ideal cylindrical mirror with perfect shape and zero roughness made of low-absorbing material neutron losses in such quasi-stationary quantum states occur via tunneling of neutrons through triangle potential barrier shown in Fig.2. The lifetime of deeply bound states is long. The lifetime of the quasi-stationary quantum states with energy close to the barrier edge is short; such neutrons tunnel rapidly into the mirror bulk. If we vary continuously the centrifugal acceleration (by means of changing the neutron velocity), we will vary  the height and the width of the  triangle barrier correspondingly and so far the lifetime of the quasi-stationary quantum states. Due to the very fast (exponential) increase of the mentioned lifetime as a function of the barrier width  we will get  a step-wise dependence of the neutron flux parallel to the mirror surface  as a function of the neutron velocity. Analogous   step-wise dependence of the total neutron flux as a function of the slit size was observed in case of the  gravitationally bound quantum states. An alternative method for observation and studying the centrifugal quantum states consists in measuring velocity distribution in the quantum states using a position-sensitive neutron detector, placed at some distance from the curved mirror. Such a method was  used as well in the experimental studies of the gravitationally bound quantum states of neutrons.

It is natural to chose the neutron velocity within the range of maximum intensity of standard neutron sources (neutron reactors or spallation sources) around $\sim 10^3$ m/s. The cylindrical mirror radius has to be equal to a few centimeters in this case that is just optimal for its production. Neutron beams with high intensity are available in many neutron centers around the world; they could be angularly and spatially collimated; time-of-flight and polarization analysis are available at standard neutron-scattering instruments. Evidently, the characteristic size of the centrifugal quantum states is much smaller than the mirror radius and the effective centrifugal acceleration could be approximated as constant with high accuracy. On the other hand, Fermi-potential of a mirror can not be considered as infinitely high. Just the opposite: quantum states energy is close to the value of Fermi-potential. Therefore in contrast to the gravitationally bound quantum states neutrons tunnel deeply into the mirror (compared to the characteristic size of the wave-functions). This phenomenon has to be taken into account. Another essential difference is related to the effects of surface roughness: as far as the characteristic scale of the centrifugal quantum states is much smaller, the roughness effects are much larger; therefore constraints for the cylindrical mirror surface are even more severe than those for flat mirrors in the gravitational experiments.
We will show rigorously in the following chapters that  the centrifugal quantum states could be described in very similar way as the gravitational quantum states although they are formed by completely different physical potentials. Large difference in characteristic scales of the quantum states in two cases requires different approaches for their experimental observation and study.

\begin{figure}
  \centering
 \includegraphics[width=90mm]{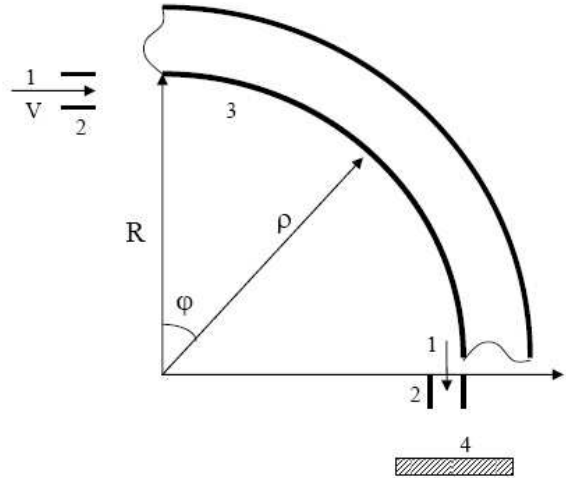}
\caption{A scheme of the neutron centrifugal  experiment. 1 -
 the classical trajectories of incoming and outcoming neutrons, 2 - the collimators, 3 - the cylindrical mirror, 4 - the detector. Cylindrical coordinates $\rho - \varphi$ used throughout the paper are shown.}\label{Sceme}
\end{figure}
\section{Formal solution}

Neutrons with given energy $E$, scattered by the curved mirror 
  obey the following
Schr\"{o}dinger equation in the cylindrical coordinates:
\begin{equation}\label{SchrEq}
\left[-\frac{\hbar^2}{2M}\left(\frac{\partial^2}{\partial\rho^2}+\frac{1}{\rho}\frac{\partial}{\partial
\rho}\right)-\frac{\hbar^2}{2M\rho^2}\frac{\partial^2}{\partial\varphi^2}+U(\rho,\varphi)-E\right]\Psi(\rho,\varphi)=0
\end{equation}
Here $M$ is the neutron mass, $\rho$  is the radial distance, measured from the center of mirror
curvature (see Fig.\ref{Sceme}), $\varphi$ is the angle and $U(\rho,\varphi)$ is the
mirror Fermi potential. We will use the following step-like
dependence for the mirror Fermi-potential:
\[
U(\rho,\varphi)=U_0
\Theta(\rho-R)\left(\Theta(\varphi)-\Theta(\varphi-\varphi_0)\right)
\]
where $R$ is the mirror curvature radius, and the angle $\varphi_0$
is determined by the mirror length $L_{mirr}$ and the mirror
curvature radius $R$ via:
\[
\varphi_0=\frac{L_{mirr}}{2\pi R}
\]
In the equation (\ref{SchrEq}) we omitted trivial dependence on z
coordinate along  the curved mirror axis. By standard
substitution $\Psi(\rho,\varphi)=\Phi(\rho,\varphi)/\sqrt{\rho}$ the
equation (\ref{SchrEq}) is transformed to the following form:
\begin{equation}
\label{SchrEq1}
\left[-\frac{\hbar^2}{2M}\left(\frac{\partial^2}{\partial\rho^2}\right)-
\frac{\hbar^2}{2M\rho^2}\left(\frac{\partial^2}{\partial\varphi^2}+\frac{1}{4}\right)+U(\rho,\varphi)-E\right]\Phi(\rho,\varphi)=0
\end{equation}

 Now the problem
is  formulated as follows.  The incoming neutron flux is known at the   curved
mirror entrance. We have to find the neutron flux  at
the exit of the mirror with the angle coordinate $\varphi_0$.
The measured  neutron current component, parallel to the mirror surface is:
\begin{equation}
\label{Current}
J(\rho,\varphi) = \frac{i\hbar}{2M\rho} \left ( \Psi(\rho,\varphi)\frac{\partial \Psi^{*}(\rho,\varphi)}{\partial \varphi}- \Psi^{*}(\rho,\varphi)\frac{\partial \Psi(\rho,\varphi)}{\partial \varphi}\right )
\end{equation}

We  start with the formal solution of the equation (\ref{SchrEq}) in
the domain $0\leq \varphi\leq \varphi_0$. We express a solution of the
equation (\ref{SchrEq}) as a series expansion in the complete set of
basis functions $\chi_{\mu }(\rho)$ \cite{Esry,Sols,Lin,Olen} :
\begin{equation}
\label{Expand} \Phi(\rho,\varphi)=\sum_{\mu} \chi_{\mu }(\rho)\left( c_{\mu}\exp(i
\mu \varphi)+d_{\mu}\exp(-i
\mu \varphi) \right)
\end{equation}
where $c_{\mu}$ and $d_{\mu}$ are the expansion coefficients.
The basis functions $\chi_{\mu }(\rho)$ are  solutions of the
following eigenvalue problem :
\begin{eqnarray}
\label{Eigenvalue}
\left[-\frac{\hbar^2}{2M}\left(\frac{\partial^2}{\partial\rho^2}\right)
+U_0 \Theta(\rho-R)-E\right]\chi_{\mu}(\rho)=-\frac{\hbar^2
(\mu^2-1/4)}{2M\rho^2}\chi_{\mu}(\rho)\\
\label{Boundary0}
\chi_{\mu}(\rho \rightarrow 0)=0\\
\label{Boundary1}
\chi_{\mu}(\rho\rightarrow \infty)=\sin(\sqrt{2M E}
\rho+\delta_{\mu})
\end{eqnarray}

Here $\hbar^2 (1/4-\mu^2)/(2M) \equiv -\hbar^2\eta^2/(2M)$ is the  eigenvalue and $\delta_{\mu}$ is the  scattering phase.  $\mu$ plays a role of the angular momentum. Let us
note that  the energy $E$ is a
fixed parameter in the equation (\ref{Eigenvalue}), while $\mu$ is the  angular momentum eigenvalue to be found.

For the above mentioned eigenvalue problem (\ref{Eigenvalue},\ref{Boundary0},\ref{Boundary1})  self-adjointness of the corresponding
Hamiltonian of radial motion is required for the completeness of the basis set $\chi_{\mu}$ \cite{PerPopov}.  One can prove  that this requirement and the boundary conditions (\ref{Boundary0},\ref{Boundary1})  are
equivalent to the following condition for the
 eigenstates \textit{phases}:
\begin{equation}\label{SelfAdj}
\delta_{\mu'}-\delta_{\mu}=\pi k
\end{equation}
where $k$ is integer. In this case, the functions $\chi_{\mu}$ are orthogonal to each other with the weight of $1/\rho^2$ on the interval $[0,\infty)$.

Note, that there is no uniqueness condition for the wave function as long as $\varphi_0<2\pi$. So far  $\mu$ is no longer an integer value in our problem.

For given positive energy $E>0$, the
values $\mu$  form  discrete spectrum of \emph{real} values if $\eta^2\geq 0$ and continuum spectrum of complex values if $\eta^2<0$.

The flux (\ref{Current}) through a band with the radial coordinates $(\rho_1,\rho_2)$  orthogonal to the mirror surface  in the mentioned basis can be expressed as:
\begin{equation}\label{CurrentInt}
F(\varphi)=\int_{\rho_1}^{\rho_2}J(\rho,\varphi) d\rho=\frac{\hbar}{M}\mathop{\rm}{Re}\sum_{\mu,\mu'}\left(\int_{\rho_1}^{\rho_2}\frac{\chi_{\mu'}^*(\rho)\chi_{\mu}(\rho)}{\rho^2}d\rho\right )\mu'
\left( c_{\mu}c^*_{\mu'} \exp (i(\mu-\mu')\varphi)-d_{\mu}d^*_{\mu'} \exp (-i(\mu-\mu')\varphi) \right)
\end{equation}
In the following we will be interested in the flux
 $F(\varphi)$ evolution   as a function of the angle $\varphi$, which
indicates the neutron density along the curved mirror. The neutron density
is ''initially'' (i.e. for $\varphi=0$) localized by the collimator near the surface
of the mirror in the band $(\rho1,\rho2)$. Due to ''dephasing'' of $\varphi$-dependent exponents in the expression (\ref{CurrentInt})
 the neutron density within the band $(\rho1,\rho2)$
decays rapidly when  $\varphi$ increases.  We will find the
rate of such a decay in the following sections. In particular we will  show that such a rate is determined by the lifetime of the quasi-stationary states
 formed by the superposition of the centrifugal potential and the Fermi potential of the mirror.

\section{Centrifugal quasi-stationary states}
To study the neutron states, localized near the mirror surface, we will expand the expression for the  centrifugal energy in the equation (\ref{Eigenvalue})  in the
vicinity of $\rho=R$. We introduce the deviation from the mirror surface $z=\rho-R$ and get the following equation
in the first order of small ratio $z/R$:
\begin{equation}\label{Rexp}
\left[-\frac{\hbar^2}{2M}\frac{\partial^2}{\partial
z^2}+U_0\Theta(z)+\hbar^2\frac{\mu_n^2-1/4}{2M
R^2}(1-2z/R)-E\right]\chi_n(z)=0
\end{equation}

We will be interested in those  solutions with different angular momenta $\mu_n$, which correspond to the states of neutron, moving parallel to the mirror surface. Such neutrons with given energy $E=Mv^2/2$  possess angular momentum $\mu_n$ close to the classical value $\mu_0=MvR/\hbar$. Let us mention that the value of $\mu_0$ is extremely high $ \mu_0\sim 5$ $ 10^8$ if  $v=1000$ m/s and
$R=2.5$ cm (parameters which can be realized in experimental setup) .
Introducing new variables $\Delta_n=\mu_0-\mu_n$, where $\Delta_n\ll \mu_0$ and $\varepsilon_n=\hbar^2 \mu_0 \Delta_n/(M R^2)$  and keeping leading terms in $\mu_0$ we get the following equation:
\begin{equation}\label{RexpLinear}
\left[-\frac{\hbar^2}{2M}\frac{\partial^2}{\partial
z^2}+U_0\Theta(z)-\frac{Mv^2}{R}z-\varepsilon_n\right]\chi_n(z)=0
\end{equation}

Let us mention that the eigenvalue $\varepsilon_n$  plays a role of energy in the above equation only formally. In fact it defines the angular momentum eigenvalue
\begin{equation}\label{AngEnergy}
\mu_n=\mu_0-\frac{\varepsilon_n M R^2}{\mu_0 \hbar^2}
\end{equation}
while the neutron energy $E$ is a fixed parameter in our problem. The value $\varepsilon_n$ can be interpreted as the radial motion energy within the above used   linear expansion of the centrifugal potential in the vicinity of the curved mirror radius $R$.

\begin{figure}
  \centering
 \includegraphics[width=115mm]{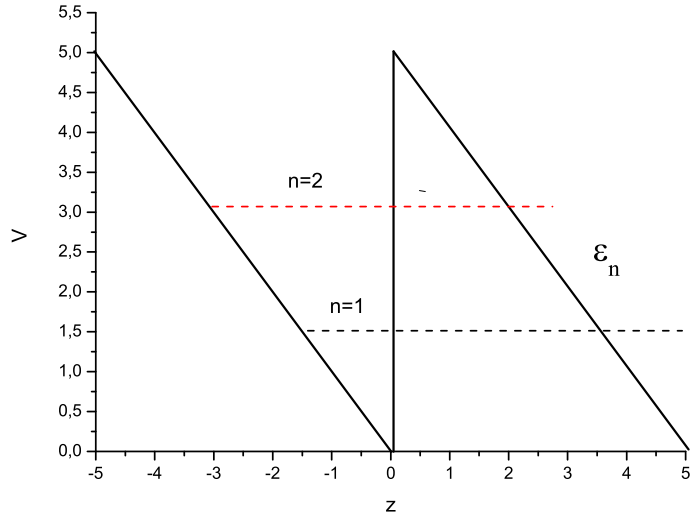}
\caption{A sketch of the effective potential in the mirror surface vicinity. The potential step at $z=0$ is equal to the mirror Fermi-potential in units $\varepsilon_0=(\hbar^2Mv^4/(2R^2))^{1/3}$. The potential slope at $z\neq 0$ is governed by the centrifugal effective acceleration $a= v^2/R$. }\label{Pot}
\end{figure}

The equation (\ref{RexpLinear}) describes the neutron motion in constant
effective  field $a=-v^2/R$ superposed with the mirror Fermi potential
$U_0\Theta(z)$.
The sketch of corresponding potential is shown in Fig.\ref{Pot}.

The regular solution of the equation (\ref{RexpLinear}) is given by the
well-known Airy function \cite{Abramowitz}:
\begin{equation}\label{Ai}
\chi_n(z)\sim \left\{\begin{array}{cll}
\mathop{\rm Ai}(z_0-z/l_0-\varepsilon_n/\varepsilon_0) & \mbox{if} & z > 0 \\
\mathop{\rm Ai}(-z/l_0-\varepsilon_n/\varepsilon_0) & \mbox{if} & z \leq 0%
\end{array}
\right.
\end{equation}
Here
\begin{equation}\label{L0}
l_0=(\hbar^2R/(2M^2v^2))^{1/3}
\end{equation}
 is the characteristic distance
scale of the problem, and
\begin{equation}\label{E0}
\varepsilon_0=(\hbar^2Mv^4/(2R^2))^{1/3}
\end{equation}
 is the
characteristic energy scale,
$z_0=U_0/\varepsilon_0$. For the typical experimental setup parameters $U_0=150$ neV,  $v=1000$ m/s and
$R=2.5$ cm
the above mentioned scales are $l_0=0.04$ $\mu$m and
$\varepsilon_0=15.3$ neV and $z_0\simeq 10$.

\begin{figure}
  \centering
 \includegraphics[width=115mm]{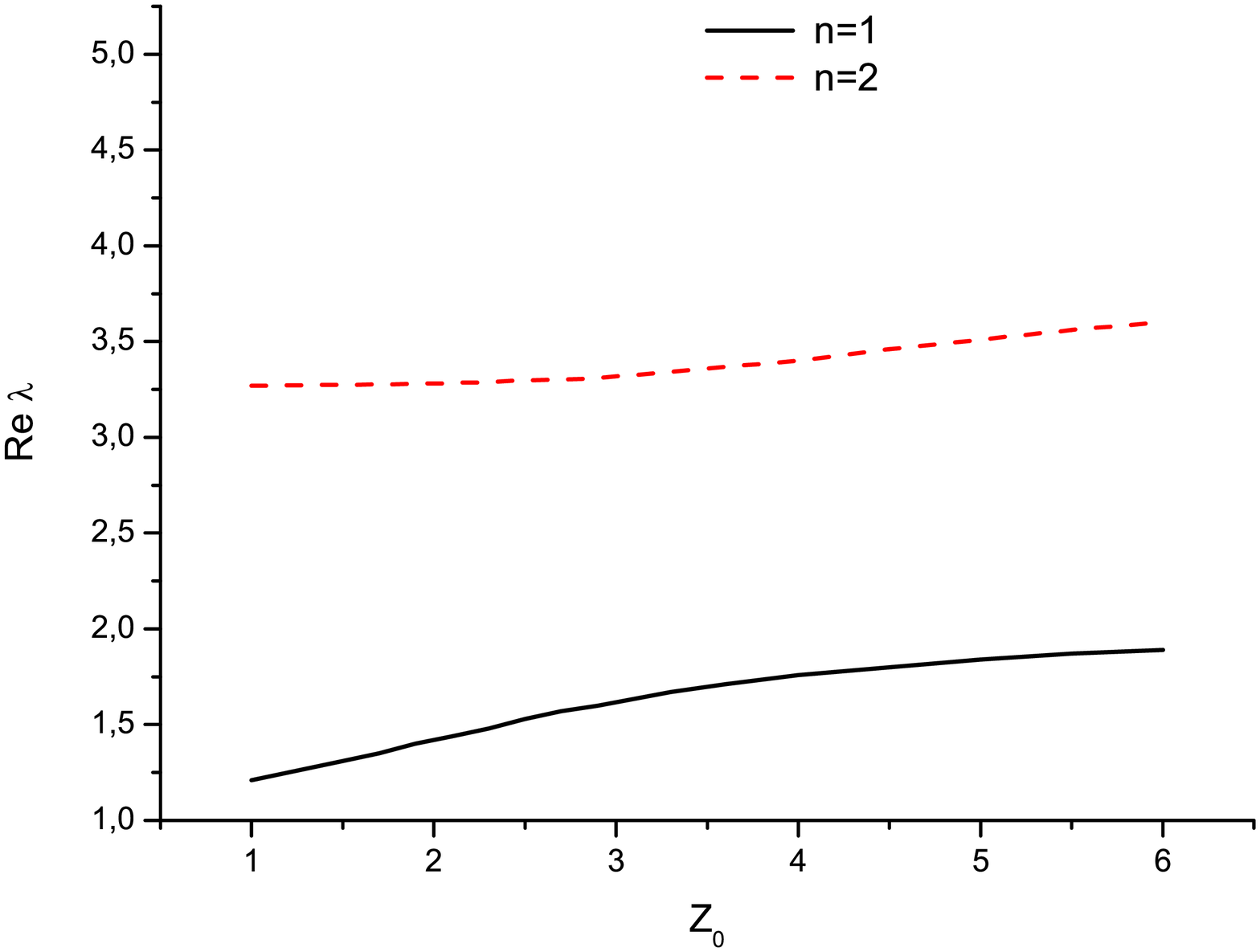}
\caption{The real part of two lowest eigen-values $\lambda$ as a function of $z_0$ obtained by numerical integration of the equation (\ref{LifeTime}). }
\label{ReLambda}
\end{figure}

\begin{figure}
  \centering
 \includegraphics[width=115mm]{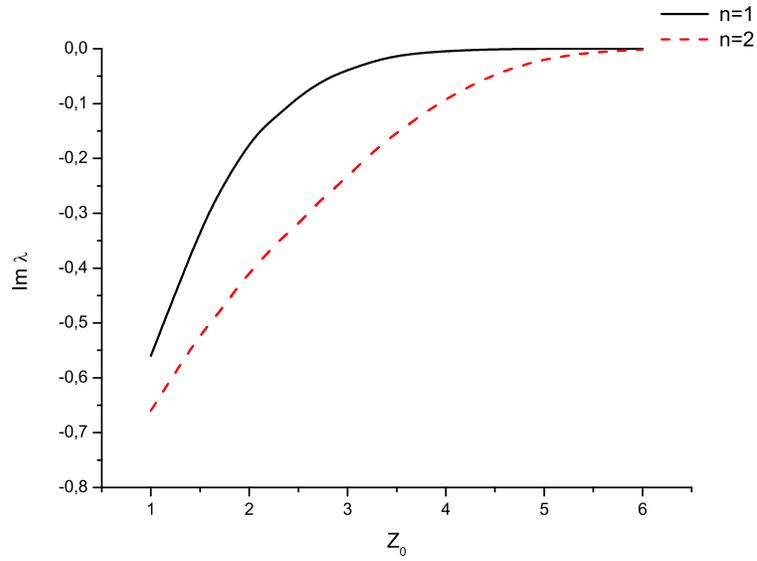}
\caption{The imaginary part of two lowest eigen-values $\lambda$ as a function of $z_0$ obtained by numerical integration of the equation (\ref{LifeTime}). }
\label{ImLambda}
\end{figure}

The  above mentioned effective potential  supports  existence of the quasi-stationary states. They
correspond to the solution of the equation (\ref{RexpLinear}) with the
outgoing wave boundary condition:
\begin{equation}\label{AiQS}
\widetilde{\chi}_n(z)\sim \left\{\begin{array}{cll}
\mathop{\rm Bi}(z_0-z/l_0-\varepsilon_n/\varepsilon_0)+i\mathop{\rm Ai}(z_0-z/l_0-\varepsilon_n/\varepsilon_0) & \mbox{if} & z > 0 \\
\mathop{\rm Ai}(-z/l_0-\varepsilon_n/\varepsilon_0) & \mbox{if} & z \leq 0%
\end{array}
\right.
\end{equation}
The complex energies of such quasi-stationary states can be found
from the matching of logarithmic derivative at $z=0$:
\begin{eqnarray}\label{enerQS}
\varepsilon_n&\equiv&\varepsilon_0 \lambda_n \\
\label{LifeTime}
\mathop{\rm Ai'}(-\lambda_n)\left(\mathop{\rm
Bi}(z_0-\lambda_n)+i\mathop{\rm
Ai}(z_0-\lambda_n)\right)&=&\mathop{\rm
Ai}(-\lambda_n)\left(\mathop{\rm Bi'}(z_0-\lambda_n)+i\mathop{\rm
Ai'}(z_0-\lambda_n)\right)
\end{eqnarray}
The real and imaginary parts of eigen-value $\lambda$, obtained by numerical solution of the equation (\ref{LifeTime}) for two lowest states are shown as a function of dimensionless variable $z_0$ in Fig.\ref{ReLambda} and Fig.\ref{ImLambda}.


One can get  semiclassical approximation for the widths of the centrifugal
quasi-stationary states  if $|\lambda_n|\gg 1$,
$z_0\gg |\lambda_n|$ \cite{PRD2}. In this case one can use the asymptotic expressions
 for the Airy functions of large argument to get the following equation:
 \begin{equation}\label{QuantCondSemicl}
 \sqrt{\frac{\lambda_n}{z_0-\lambda_n}}\tan(\frac{2}{3}\lambda_n^{3/2}-\pi/4)=1-2i\exp(-4/3(z_0-\lambda_n)^{3/2})
 \end{equation}
 Also one can get semiclassical approximation for  the width $\Gamma_n$ and $\lambda_n$ from the above expression, valid for large $n$:
\begin{eqnarray}\label{lambda}
\lambda_n&\simeq&(\frac{3}{4}\pi (2n - 1/2))^{2/3}-\sqrt{\frac{(\frac{3}{4}\pi (2n - 1/2))^{2/3}}{z_0-(\frac{3}{4}\pi (2n - 1/2))^{2/3}}}\\
\label{Gamma}
\Gamma_n&\simeq&4\varepsilon_0 \frac{\sqrt{z_0-\lambda_n}}{z_0} \exp(-4/3(z_0-\lambda_n)^{3/2})
\end{eqnarray}
In the above expressions $n=1,2,...$ is an integer number.
The angular momentum eigenvalue, corresponding to the complex energy $\varepsilon_n$ of the quasi-stationary states,
obtains  positive imaginary part, according to (\ref{AngEnergy}):
\begin{equation}\label{ImMoment}
\mathop{\rm Im}\mu_n=\frac{\Gamma_n R}{2 \hbar v}
\end{equation}

\begin{figure}
  \centering
 \includegraphics[width=125mm]{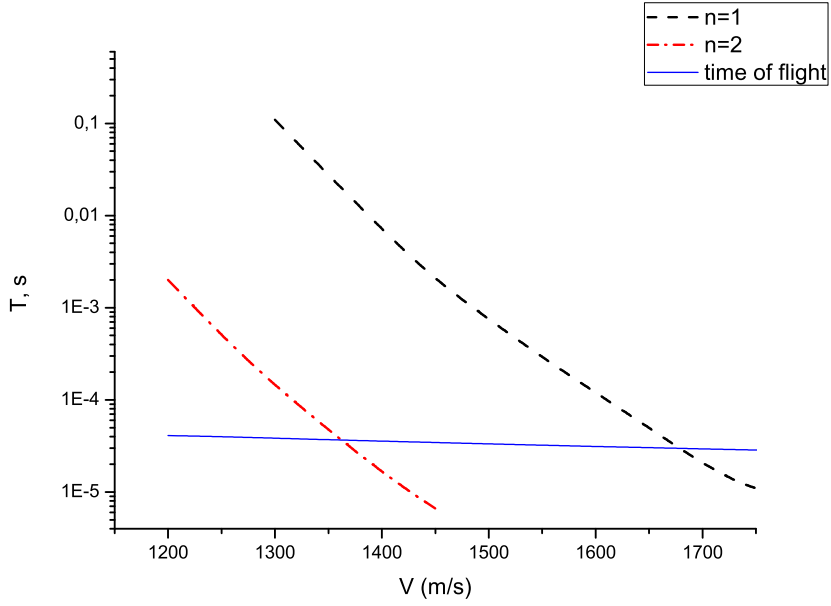}
\caption{The life-time of the two lowest neutron centrifugal quasi-stationary
states is shown as a function of the neutron velocity. The mirror  curvature radius equals R=2.5 cm, the mirror length is 5 cm, and the mirror Fermi potential is $U_0=150$ neV.
 Nearly horizontal solid line
indicates the  time of flight
along the curved mirror.}\label{LTime}
\end{figure}

\begin{figure}
  \centering
 \includegraphics[width=125mm]{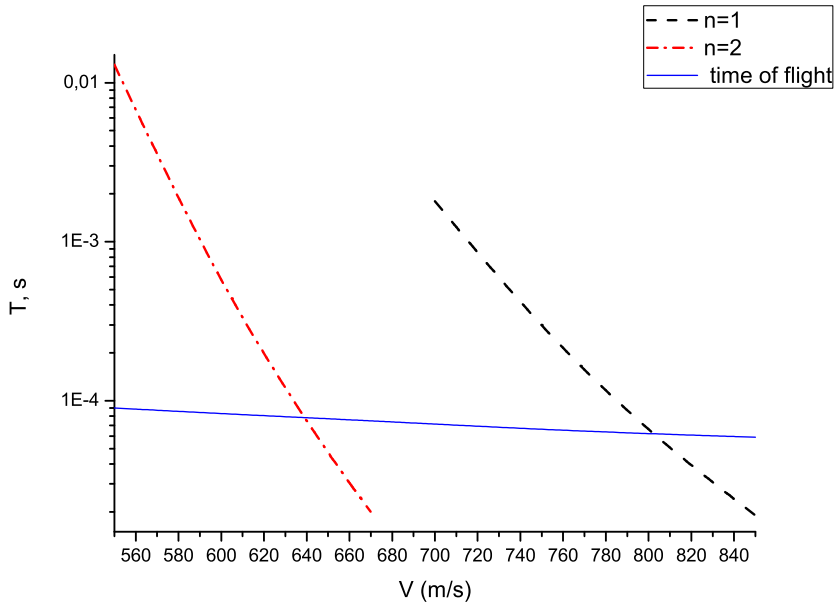}
\caption{The life-time of the two lowest neutron centrifugal quasi-stationary
states is shown as a function of the neutron velocity. The mirror  curvature radius equals R=2.5 cm, mirror length is 5 cm, and mirror Fermi potential is $U_0=54$ neV.
 Nearly horizontal solid line
indicates the  time of flight
along the curved mirror.}\label{LTime2}
\end{figure}

The energy and the width of the quasi-stationary states
depends strongly on the centrifugal acceleration $|a|=v^2/R$. Small acceleration
$a$ results in broad barrier, which separates the states in the
effective well from continuum. Indeed, $z_0=U_0/\varepsilon_0=U_0[(2R^2)/(\hbar^2 M v^4)]^{1/3}$ increases if $v$ decreases. The widths of the quasi-stationary states  decrease exponentially as it is seen from the expression (\ref{Gamma}).
  Besides that, the effective well becomes broader and new quasi-stationary states appear with decreasing of $a$ (in
analogy with appearance of new bound states with increasing the size
of the well). The equation (\ref{QuantCondSemicl}) enables us to  estimate  the critical values of the neutron velocity $v_c$, which correspond to the appearance of  new states in the effective well:
\[z_0=\lambda_n^0=(3/2\pi (n - 3/4))^{2/3}\]
Taking into account that $z_0=U_0/\varepsilon_0$ we conclude:
\begin{equation}\label{VcSemi}
v_c^n=\left[\frac{U_0^3}{(3/2\pi (n - 3/4))^{2}}\frac{2R^2}{\hbar^2M}\right]^{1/4}
\end{equation}
The accuracy of the above approximation increases with $n$.
 The lifetime of the two lowest quasi-stationary quantum states as a
function of the neutron velocity, obtained from solving the equation (\ref{LifeTime}) is shown in Fig.\ref{LTime} for the mirror with the Fermi potential $U_0=150$ neV (sapphire) and in Fig.\ref{LTime2} for the mirror with the Fermi potential $U_0=54 $ neV (silicium). The  critical velocity values  scales with the Fermi potential as $v_c\sim U_0^{3/4}$.


 The above mentioned
quasi-stationary states play essential role in
neutron density  evolution near the mirror surface as a function of $\varphi$.  We will show that under certain conditions the expansion (\ref{CurrentInt}) can be substituted by a few effective terms, corresponding to the contribution of quasi-stationary states. To clarify the role of quasi-stationary states it would be more convenient to use  time-dependent formalism.
\section{Time-dependent approach}
Let us return to the expansion (\ref{CurrentInt}) for the neutron current through the band of dimension $h=\rho_2-\rho_1\ll R$, orthogonal to the mirror surface. Taking into account very large values of the angular momenta of the neutron "near-surface" states ($\mu\sim \mu_0\simeq 5$ $10^8$), we can use the
semi-classical character of motion along the $\varphi$ variable. It is
 characterized by contribution of fast oscillating exponents $\exp(i \mu \varphi) $.
 This
enables us to treat $\varphi$ as a classical variable. Namely we will
assume that neutrons follow classical "trajectory" along $\varphi$:
\[\varphi=\omega t=\frac{vt}{R}\]
The evolution along $\varphi$ is then substituted by the  evolution  of the time-dependent wave-function. Taking into account the relation between the angular momentum and the energy eigenvalues (\ref{AngEnergy}) we come to the following expression:
\begin{equation}\label{CurrentIntTime}
F(t)=\frac{v}{R}\mathop{\rm}{Re}\sum_{n,n'}\left(\int_{\rho_1}^{\rho_2}\chi_{n'}^*(\rho)\chi_{n}(\rho)d\rho\right )
\left( c_{n}c^*_{n'} \exp (i(\varepsilon_n'-\varepsilon_n)t/\hbar) \right)
\end{equation}
with functions $\chi_n(\rho)$ being solutions of (\ref{RexpLinear}).
 In the above expression we  neglected  back-scattering and put coefficients $d_n=0$. Also we took into account that the size of the band $\rho_2-\rho_1=h \ll R$, where  the neutron density is measured.

 Thus we arrive to the problem of the time evolution (instead of  the angular variable $\varphi$ evolution) of initially localized wave-packet
 in the band with radial dimension $h=\rho2-\rho1$, which moves in the effective homogeneous field $a=v^2/R$. As it was shown in \cite{BZP} the integration over energies  in (\ref{CurrentIntTime}) results in
two  terms. One term reflects the existence of
$S$-matrix poles in the complex energy plain, which are situated
close to the real axis and correspond to the complex energies  of
quasi-stationary states. So far this term describes the decay of the
quasi-stationary states  and the characteristic time
scale is given by the corresponding widths $\tau_n=\hbar/\Gamma_n$.

The second term reflects the non-resonant contribution of all other
energies (which do not match with energies of the quasi-stationary
states).  The
characteristic time  $\tau_{cl}=\sqrt{2hR/v^2}$ for such neutrons is equal to the
classical time of passage of distance $h$ with constant
acceleration $v^2/R$. This time of passage is much smaller than the
time, which the neutron spends in the quasi-stationary states $\tau_{cl}\ll
\tau_n$.

For the times $\tau_{cl}\ll t\leq \tau_n$ the
quasi-stationary states contribution is dominant. This enables us to
neglect non-resonant contribution in the expansion (\ref{CurrentIntTime}) and
to take into account only the quasi-stationary states contribution.
\begin{equation}\label{QSExp}
F(t)\approx \frac{v}{R} \sum_{n'} |C_{n'}|^2\exp(-\Gamma_{n'}t)
\end{equation}
Here $n'$ indicates the quasi-stationary state number, $|C_{n'}|^2$ is the initial population of a given quasi-stationary state.

\begin{figure}
  \centering
 \includegraphics[width=125mm]{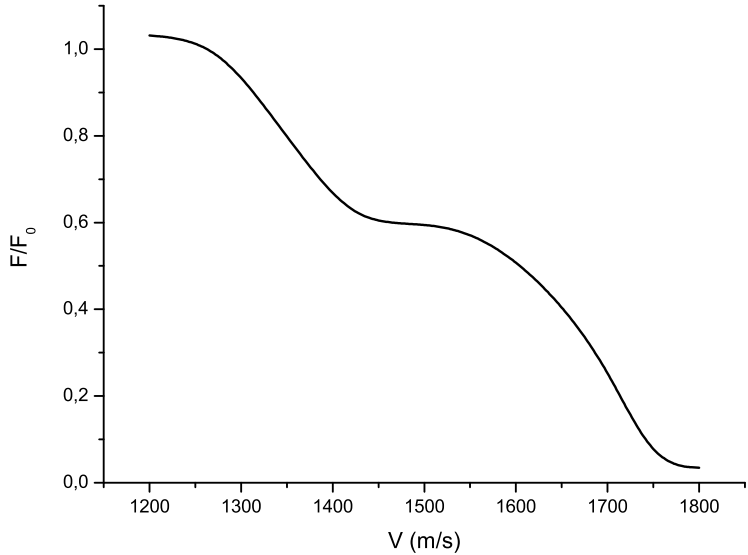}
\caption{The relative flux of neutrons, deflected by the curved mirror
as a function of the neutron velocity. $F_0$ is the flux
calculated at $v=1200$ m/s. The mirror curvature radius is $R=2.5$
cm, the mirror length is $L_{mirr}=5$ cm and the mirror Fermi potential
$U_0=150$ neV. }\label{DeflFactor}
\end{figure}

The sharp increase in the quasi-stationary states lifetime
(\ref{Gamma}) with decreasing  the velocity below $v_n$ (\ref{VcSemi})
can be used for experimental observation of such states. Indeed,
when the neutron velocity decreases  the
contribution of new quasi-stationary states increases rapidly. This
results to the step-like dependence of the deflected
neutron flux as a function of $v$.  There are no
quasi-stationary states for $v\gg v_c^1$ and therefor all neutrons traverse the mirror without
being deflected. There are many
quasi-stationary states in the opposite limit $v\ll v_c^1$ and therefor we deal with the classical reflection from the
curved mirror. In Fig.\ref{DeflFactor} the flux of
deflected neutrons is shown  as a function of the neutron velocity.
Under the assumptions made above the problem of deflection of
cold neutrons ($v\sim 10^3$ m/s) by the curved mirror is
analogous to the problem of the passage of ultra-cold neutrons
through the slit between a horizontal mirror and an absorber in the
presence of the Earth's gravitational field, studied in details in
\cite{Nesv1,Nesv2,PRD1,EPJC,PRD2,MeyNesv1,MeyNesv2,West}. In the cited experiments the spatial
density of neutrons in the gravitational states was scanned by changing
the position of the  absorber above the mirror. In case of neutron motion
along the  curved mirror surface  the initial
velocity variation results in changing the spatial dimension of the effective
well, which bounds the neutron near the surface, ensuring the
''scanning'' of the quasi-stationary states.

The experimental observation of the centrifugal states of neutrons could
be however complicated by the diffuse scattering of neutrons from the the mirror surface roughness.
Below we will estimate the additional broadening of the
centrifugal states due to the scattering on rough surface.

\section {Effect of roughness}
The effect of roughness consists in transferring  the high neutron velocity
parallel to the mirror surface  into velocity component normal to the surface.
As a result the quasi-stationary centrifugal states acquire
additional "ionization" width. The detailed theory of neutron
rough-surface interaction can be found in \cite{MeyNesv1,MeyNesv2}. To obtain a
simple estimation of such a width we will follow the method,
developed in \cite{PRD2}. Namely, in the frame related to the
neutron the mirror roughness  appears as a time-dependent
variation of the mirror position. We will start by treating
a simple case of harmonic dependence:
\[
U(z)=U_0\Theta(z+b_r\sin(\omega_r t))
\]
Here $b_r$ is the roughness amplitude and $\omega_r$ is the roughness frequency, which can be related to the angular velocity of
neutrons $\omega$, the mirror curvature radius $R$ and the characteristic
length of roughness $l_r$ via $\omega_r=\omega R/l_r$. Then the equation
describing the evolution of initially localized wave-packet gets the time-dependent right-hand side:
\begin{equation}\label{TDR}
i\hbar\frac{d\Psi(z,t)}{dt}=\left[-\frac{\hbar^2}{2M}\frac{\partial^2}{\partial
z^2}+U_0\Theta(z-b_r\sin(\omega_r
t))-\frac{Mv^2}{R}z\right]\Psi(z,t)
\end{equation}

A solution of such an equation could be expanded in the set of
eigen-functions of right-hand side Hamiltonian, taken at instant
$t$:
\begin{equation}\label{eigenTD}
\left[-\frac{\hbar^2}{2M}\frac{\partial^2}{\partial
z^2}+U_0\Theta(z-b_r\sin(\omega_r
t))-\frac{Mv^2}{R}z-\varepsilon_n(t)\right]u_n(z,t)=0
\end{equation}

The corresponding expansion is:
\begin{equation} \label{ExpR}
\Psi(z,t)=\sum_n C_n(t)u_n(z,t)\exp(-i\int_0^t
\varepsilon(\tau)/\hbar d\tau)
\end{equation}

Substitution of (\ref{ExpR}) into (\ref{TDR}) yields in the coupled
equation system for time-dependent amplitudes $C_n(t)$:
\begin{equation}\label{CnT}
\frac{dC_n(t)}{dt}=-\sum_k\langle u_n|\frac{d}{dt}|u_k\rangle C_k(t)
\exp(-i\omega_{nk}(t))
\end{equation}
Here $\omega_{nk}(t)= \int_0^t
(\varepsilon_k(\tau)-\varepsilon_n(\tau))/ \hbar d\tau$. It follows
directly from (\ref{eigenTD}), that
\[\langle u_n|\frac{d}{dt}|u_k\rangle=\frac{\langle u_n|\frac{dU}{dt}|u_k\rangle}{\varepsilon_n-\varepsilon_k}
\]
 In the following we will consider
roughness small enough, so that $U(z,t)\approx
U_0\Theta(z)+U_0b_r\omega_r \cos(\omega_r t)\delta(z)$. The coupling
matrix elements are:
\[
\langle u_n|\frac{dU(z,t)}{dt}|u_k\rangle=b_rU_0\omega_r
\cos(\omega_r t)u_n(0,t)u_k(0,t)\approx b_rU_0\omega_r \cos(\omega_r
t)u_n(0,t=0)u_k(0,t=0)
\]
Using an analog of the Fermi "golden rule" we get  the following
expression for ionization
probability of centrifugal state $n$ per unit of time:
\begin{equation} \label{FGR}
P_{ion}=\frac{2\pi b_r^2U_0^2
|u_n(0,t=0)u_f(0,t=0)|^2}{\hbar}\delta(\varepsilon_n+\hbar
\omega_r-E_f)dk_f
\end{equation}
Here index $f$ labels the eigen-state of the final state of
continuum spectrum with energy $E_f$ and wave-number $k_f$. We assume
 that $E_f\gg \varepsilon_n$ for the neutron
velocity $v\sim 10^3$ m/s and for realistic roughness parameters. This
enables us to use the free-wave
expression $u_f=1/\sqrt{2\pi}\exp(ik_f z)$ for the final state wave-function. Taking into account the
explicit form of $u_n$ given by (\ref{AiQS}) and its semiclassical asymptotic we  get
simple estimation for $P_{ion}$ of the n-th quasi-stationary state:
\begin{equation}\label{PionSC}
P^n_{ion}\approx \frac{b_r^2 U_0^2 }{\hbar^2 R l_0 (z_0-\lambda_n)\sqrt{2M
E_f}}
\end{equation}
Taking into account the explicit expressions for the characteristic length (\ref{L0}) and energy (\ref{E0}) scales of the problem, we get for the case $z_0\gg \lambda_n$:
\begin{equation}\label{Pion}
P_{ion}\approx \frac{b_r^2 U_0 v^2 M^2}{\hbar^2 R \sqrt{2M
E_f}}
\end{equation}

To get the ionization width of the centrifugal state one should
integrate the obtained probability with the spectral function of
roughness $f(\omega)$, which provides the square of
roughness amplitude as a function frequency:
\begin{equation} \label{GammaIon}
\Gamma_i=\hbar\int_0^{\infty}\frac{b_r^2 f(\omega)U_0 v^2
M^2}{\hbar^2 R \sqrt{2M (\varepsilon_n+\hbar
\omega)}}d\omega=\frac{\overline{b_r^2} U_0 v^2 M^2}{\hbar R
\sqrt{2M \overline{E_f}}}
\end{equation}
 Here $\overline{b_r^2}$ is the mean square roughness, and $\overline{E_f}$ is the mean ionization energy in the sense defined.

\begin{figure}
  \centering
 \includegraphics[width=115mm]{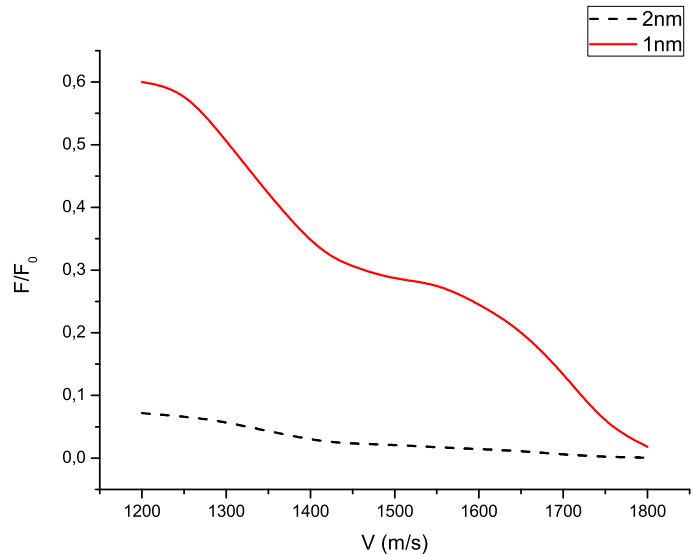}
\caption{The relative flux of neutrons, deflected by the curved sapphire mirror
as a function of the neutron velocity. $F_0$ is the flux
calculated at $v=1200$ m/s and zero roughness. The mirror curvature radius equals $R=2.5$
cm, the mirror length is $L_{mirr}=5$ cm, the mirror Fermi potential is
$U_0=150$ neV . Solid line corresponds to the roughness amplitude $b_r=1$ nm, dashed line corresponds to the roughness amplitude $b_r=2$ nm and the
roughness length
$l_r=1$ $\mu m$.}\label{Rough1}
\end{figure}

\begin{figure}
  \centering
 \includegraphics[width=115mm]{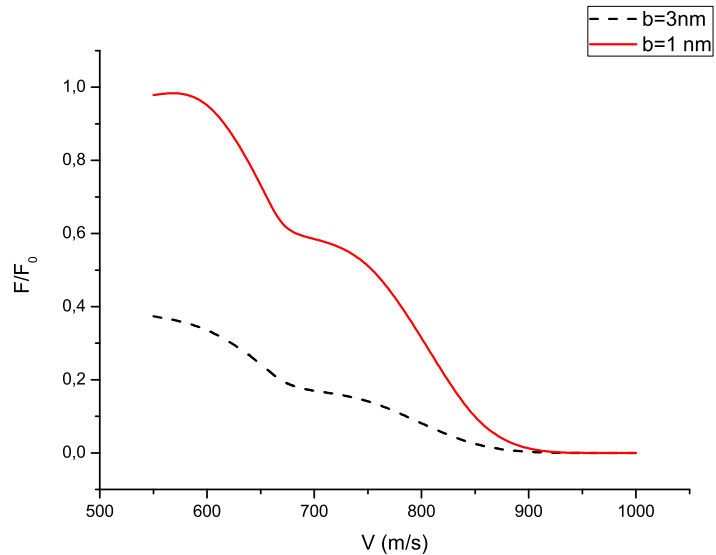}
\caption{The relative flux of neutrons, deflected by the curved silicon mirror
as a function of the neutron velocity. $F_0$ is the flux
calculated at $v=500$ m/s and zero roughness. The mirror curvature radius equals $R=2.5$
cm, the mirror length is $L_{mirr}=5$ cm, the mirror Fermi potential
$U_0=54$ neV , solid line corresponds to roughness amplitude $b_r=1$ nm, dashed line corresponds to roughness amplitude $b_r=3$ nm,
roughness length
$l_r=1$ $\mu m$.}\label{Rough2}
\end{figure}

  An important feature of the obtained result is the square
dependence of ionization width on the neutron velocity and the roughness amplitude (for the
case of small amplitudes, studied above). It constraints severely
the roughness amplitudes acceptable for observation the centrifugal states. In Fig.\ref{Rough1} we plot the neutron flux,
deflected by the curved sapphire mirror in the presence of roughness with the amplitudes $b_r=1$ nm and $b_r=2$ nm. Fig.\ref{Rough2} demonstrates the neutron flux,
deflected by the curved silicon mirror in the presence of roughness with the  amplitudes $b_r=1$ nm and $b_r=3$ nm. Thus the effect of roughness is reduced if the Fermi potential is low. Indeed, according to (\ref{VcSemi}) and (\ref{GammaIon}) we expect the following scaling law:
\[ P_{ion}\sim U_0^{17/8} \]
 So the roughness amplitude of a sapphire mirror surface should
 be smaller than $1$ nm  (and $4$ nm for silicon mirror) to allow observation of the
 centrifugal states.
 \section{Conclusions}
 We proposed a method for observation of the quasi-stationary states of
 neutrons, localized near a curved mirror surface. The
 effective bounding well is formed by a superposition
 of the centrifugal potential and  the mirror Fermi potential. Reduction of the initial neutron velocity results in the spatial size increase of
   such a centrifugal trap which, in its turn, results in
 appearance   of the  quasi-stationary states in the spectrum of the
 system. This could be observed via step-like dependence of the deflected neutron flux.
  We show that several centrifugal states can be
 observed for instance with a sapphire mirror (Fermi potential $U_0=150$ neV), with the curvature radius $R=2.5$ cm, the length $L_{mirr}\approx 5$ cm and the surface roughness
 amplitude $< 1$ nm.
The critical velocities corresponding to the steps in the deflected
flux are $v_1=1700$ m/s and $v_2=1350$ m/s.  The characteristic
spatial dimension of the mentioned centrifugal states
 is  $l_0\approx 0.04$ $\mu$m. In case of a silicon mirror with the same shape (Fermi potential $U_0=54$ neV) the corresponding  critical velocities values are $v_1=810$ m/s and $v_2=650$ m/s.
 Such neutron states could provide  a promising  tool for studding different types of  neutron-matter interactions with the characteristic range of a
 few tens nm.
 \section{Acknowledgement}
We are grateful to our colleagues from GRANIT collaboration and participants of GRANIT-2006 Workshop for stimulating discussions and ANR (Agence Nationale de la Recherche, France) for partial support of this work.

\bibliographystyle{unsrt}
\bibliography{curvedmirr}

\end{document}